\documentstyle[11pt,newpasp,epsf,epsfig,twoside]{article}
\def\edcomment#1{\iffalse\marginpar{\raggedright\sl#1\/}\else\relax\fi}
\reversemarginpar

\begin{document}
\title{Spiral waves in accretion discs}

\author{H.M.J. Boffin} 

\affil{Royal Observatory of Belgium, 3 av. Circulaire, 1180 Brussels}

\author{D. Steeghs} 

\affil{Physics \& Astronomy, Southampton University, Southampton SO17 1BJ, UK}

\begin{abstract}
In the first part of this article, we review the observational evidence
for spirals in the accretion discs of cataclysmic variables. It is shown that with the increasing amount of data available, spirals appear to be
an omnipresent feature of accretion discs in outburst. Spirals seem to live
until decline that is, for several tens of orbital periods. 
We then study the formation of spiral shocks from a theoretical side, using
the results of various numerical simulations. We make a comparison 
between observations and theory and briefly discuss the implications of the 
presence of spirals in the discs of cataclysmic variables.
\end{abstract}

\section{Introduction}

In close binaries,  it is  the   detailed  process  of  angular momentum   transport  that
determines the structure of the accretion disc and therefore the rate
at which gas,  supplied from the mass  donor, is actually  accreted by
the   compact   object.  Although  so fundamental   to  the process of
accretion   through discs,   our  understanding of   angular momentum
dispersal  is   very limited.   We  can  roughly  divide the  possible
physical  mechanisms that  may provide  the required angular  momentum
transport  into  two classes.  Those that  work  on  a local scale and
exchange angular momentum among neighbouring  parcels in the disc, and
those  that  rely on global, large    scale structures in  the disc.  The local
processes are commonly referred to  as `viscous processes' though
it  is clear that the  molecular  viscosity of  the accretion  disc
material itself is many orders of magnitudes too small.
Viscous interaction in the sheared Keplerian disc allows some material
to  spiral inwards, losing  angular momentum, while excess momentum is
carried   outwards by   other parts  of    the  flow.  In the   famous
$\alpha$-parameterisation of  Shakura  \&  Sunyaev  (1973),  this
viscosity was replaced   by  a single  dimensionless  constant,  which
allows one to solve the structure equations for thin, viscously heated
accretion discs.

A  very  different way  of  transporting the   angular momentum is via
density waves in the disc.  In self-gravitating  discs, the ability of
density  waves to transport angular  momentum  is a direct result from
purely gravitational interaction between  the wave and the  surrounding
disc material.   Waves  of this type  can
still  transport   angular  momentum  in  the absence  of self gravity,
provided  some mechanism exists   that exchanges momentum between  the
wave  and the fluid. Sawada, Matsuda, \& Hachisu  (1986) conducted numerical
simulations of mass transfer via Roche lobe overflow  of inviscid, non self-gravitating discs, and 
witnessed the
development of strong spiral shocks in the disc which were responsible
for  the bulk   of the   angular  momentum  transport throughout the flow. Such trailing spiral patterns are the  natural result of a tidal
deformation of the disc that is  sheared into a  spiral pattern by the
(near) Keplerian rotation profile of the disc material (Savonije, Papaloizou, \& Lin 1994).

Although predicted   in  the 80s, observational evidence   for spirals
relies on  the  ability to  spatially resolve  the accretion  discs in
interacting binaries.  Indirect imaging methods   are thus required to
search for such global disc asymmetries.   Since the wave pattern is a
co-rotating structure close to  the orbital plane,  
Doppler tomography
of strong emission lines is the ideal tool at hand. Although providing
an image of the line emission distribution in  velocity space, and not
spatial   coordinates, 
spirals should be    readily identified as they
maintain their spiral shape in the velocity coordinate frame.

The first convincing evidence of spirals was found in 1997, in the accretion disc
of the eclipsing dwarf nova IP Peg. Since then, such waves have been confirmed in several
outburst events of IP Peg as well as in other cataclysmic variables.
In the next section, we will review the current observational status. A more detailed review can be found in Steeghs (2001).

\section{Observational overview}
\subsection{IP Peg: when the story begun}
When a Doppler map was constructed of the hydrogen and helium emission
from IP Pegasi, just
after rise  to one  of its  outbursts, a surprising emission pattern was found
(Steeghs, Harlaftis, \& Horne, 1997, 1998). 
The accretion disc was far from symmetric but instead disc emission
was dominated by a two armed pattern in the lower left and upper right
quadrants of  the  Doppler map.  The spiral  arm
velocities ranged between 500 and 700 km/s,  corresponding to the outer
regions  of the accretion disc.   The  emissivity contrast between the
spirals and other parts of the disc, was about a  factor of
 $\sim$3 for
H$\alpha$,  and $\sim$5 in the case  of HeI 6678  emission. There was no
evidence for line emission from the bright spot. The spiral arm in
the lower quadrant, closest to the secondary, extended over an angle of
$\sim$100$^{\circ}$, and  was   weaker than the  arm in   the  opposite
quadrant. Strong  emission from the irradiated  secondary star was also
present, producing the prominent S-wave at low velocities.

A week later, during the same outburst, more spectroscopy was
 obtained
before  the decline of  the outburst  had started (Steeghs et al. 1996).   
The Doppler image showed   that the   spiral pattern  persists
throughout the   outburst,   and the  secondary   star  made   a
considerably larger contribution (from 6\% to  10\%) to the line flux.
The  arm  in  the  upper  right quadrant  was still   stronger, and the
location  of the arms had not  changed, although the  upper right arm
appeared shorter.  

In a  different outburst Harlaftis et al. (1999) secured a  whole orbit
of   IP Pegasi, three days after outburst.
The HeII emission line  at 4686\AA~ was observed
to provide a comparison with the previously observed H$\alpha$ and HeI
emission patterns.  The trailed spectrogram of HeII,   again showed the  familiar  behaviour of the double
peaks from the  disc, leading to  a  two armed  spiral in  the Doppler
tomogram. A  very   similar emission pattern from  the   disc was also
present in the nearby Bowen blend consisting of CIII/NIII emission and
the Helium  I line at 4471\AA. The two spirals could be
traced for  almost 180 degrees, and  the  velocity of the  arms varied
from 495 km/s to 780 km/s. The  widths of  the  arms were significant, and  even
changed as a function of azimuth. It is thus an intrinsically
broad  feature: the arms cover a substantial part  of the disc,
with the strongest emission from the
outer regions. The dynamics of the majority of the accretion disc material are thus affected by the presence of these 
spirals.

Morales-Rueda, Marsh \& Billington (2000)  present spectroscopy 
of the dwarf nova IP~Pegasi taken during two
consecutive nights, 5 and 6 days after the start of the August 1994 outburst.
Doppler maps show marked spiral
structure in the accretion disc. The spiral shocks are present on
both nights with no diminution in strength from one night to the
next.  

The detections by Steeghs et al. (1997)
and Harlaftis et al. (1999) were made $1.5$ and 3 days after the start
of the August 1993 and the November 1996 outbursts respectively,
whereas 
the data presented by Morales-Rueda et al. (2000)  were
taken 5 and 6 days after the outburst had started. 
The spectroscopic data taken 8 days after the start of the August
1993 outburst by Steeghs et al. (1996) also hints at their presence.
This indicates that the
spiral shocks are long-lived structures, with time-scale of the order of days instead of hours.  

During the 1999 outburst of IP Peg, spirals were again observed (Steeghs \&
Boffin, in preparation). One can then conclude that 
Doppler tomography of IP Pegasi  during outburst invariably shows
the presence of spiral shaped  disc asymmetries.  The two armed spiral
dominates the disc emission from the start of the outburst maximum and
persists for at least 8 days, corresponding to about 50 binary orbits.
The spirals  are present  in a  range of  emission lines  from neutral
hydrogen   to ionised helium,  the  latter   indicating  that the  gas
concerned has to be hot, although it is not yet clear if we are looking at
direct emission from  the  shock, or  recombination  emission from the
spiral arms.  The asymmetry between the two  arms that was observed in
the  discovery data,  is also present   at  other epochs and in  other
lines.  
Moreover, as we will now see, IP Peg is no longer the only system with evidence for distortions of a spiral nature.

\subsection{Disc asymmetries in other systems}
The eclipsing dwarf nova EX\,Dra,
with a 5-hour orbit,
was the  second  object to show  similar  disc
behaviour. Joergens, Spruit, \& Rutten (2000)  present  Doppler images  of EX  Dra during its July 1996
outburst,  and the  similarity of the Doppler maps  with  those of IP
Pegasi  is obvious.
The disc seen in the HeI line reconstructed by Doppler tomography
shows a clear two-armed spiral pattern pointing to spiral shocks in the disc.
The Balmer and HeII maps also give evidence for 
the presence of spirals. 
The asymmetry of the spirals in the EX\,Dra disc  
is smaller than in the IP\,Peg observations
perhaps indicating that the shocks are weaker.
The spiral arm in the upper right is stronger in intensity as well as in
asymmetry than that one in the lower left. This pattern is also seen in the 
Doppler maps of IP\,Peg. 

Time resolved spectroscopic observations of U Gem during its March
2000 outburst show strong spiral shocks in the accretion disc 
(Groot 2001). 
The HeII 4686 most clearly shows the spiral arms during outburst. 
During the plateau at maximum brightness, 9 days into the outburst,
 the spiral shocks contribute
$\sim$14\% to the total HeII flux. The two arms of the spiral show a
distinctly different evolution during the outburst and decline, which
indicates an asymmetric evolution in the disc. 
Before outburst, the trailed spectrum shows a clear single
S-wave which is identified as arising from the hot spot,
located at a position of 0.55
R$_{\rm L_1}$. 
The upper arm is stronger than the lower one at the end of the
maximum brightness plateau but not during the decline
of the outburst.
Nine days after maximum brightness, the slope 
in the radial velocity - orbital phase plane of the lower arm of the spiral
is --2600$\pm$700 \mbox{km\,s$^{-1}$\,orbit$^{-1}$} and for the upper arm this 
is --1300$\pm$200 \mbox{km\,s$^{-1}$\,orbit$^{-1}$}. One day later, these numbers are
--1740$\pm$400 \mbox{km\,s$^{-1}$\,orbit$^{-1}$} and
--1100$\pm$200 \mbox{km\,s$^{-1}$\,orbit$^{-1}$} for the lower and upper arms, respectively,
while during the decline, they become, respectively, 250$\pm$140 \mbox{km\,s$^{-1}$\,orbit$^{-1}$} 
and 900$\pm$170 \mbox{km\,s$^{-1}$\,orbit$^{-1}$}.
Thus, the decrease of the slope of the lower arm 
is much more rapid and pronounced than for the upper one.

\begin{figure}
\plotone{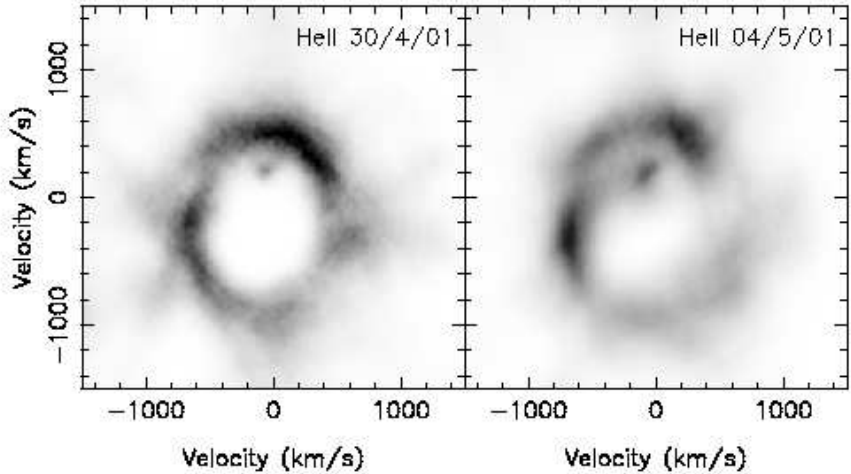}
\caption{
\label{fig:ugem}
Doppler maps of U Gem at the start of the outburst (left) and 4 days
into outburst (right). The presence of spirals which persist into the outburst is
very clear. From Steeghs et al. (in preparation). 
}
\end{figure}

As stated by Groot (2001), a possible evolutionary scenario for the spiral shocks one can derive
from the comparison of these three systems is that the shocks appear
immediately when the outburst starts (IP Peg; Steeghs et al. 1997),
but not just prior to outburst (U Gem in quiescence; Groot 2001), grow in strength when
the outburst reaches its maximum magnitude (Harlaftis et al. 1999 and
Joergens et al. 2000), and continue to gain in strength, or at least
remain constant, during the plateau phase characteristic of many DN
outbursts (Steeghs et al. 1996, Groot 2001), and then fade  
during outburst decline (Groot 2001).

In another outburst of U Gem, Steeghs, Morales-Rueda, \& Marsh (in preparation)  
also observed clear spirals in the HeII at the start of the outburst, which 
persisted during the maximum brightness	plateau, 4 to 5 days after outburst (Fig.~\ref{fig:ugem}). 
On the other hand, during decline, 9 days after outburst, the Balmer emission reveals a relatively circular disc.

Indication for the presence of spiral patterns in the stars SS Cyg and V347 Pup
are also discussed in Steeghs (2001). Very recently, prominent spiral arms have also been observed in the early phases of the 2001 WZ Sge
outburst (Kuulkers et al., this volume).

We shall now turn to the theory and see if such spirals, apparently ubiquitous in the discs of cataclysmic variables, can be explained.

\begin{figure}
\plotone{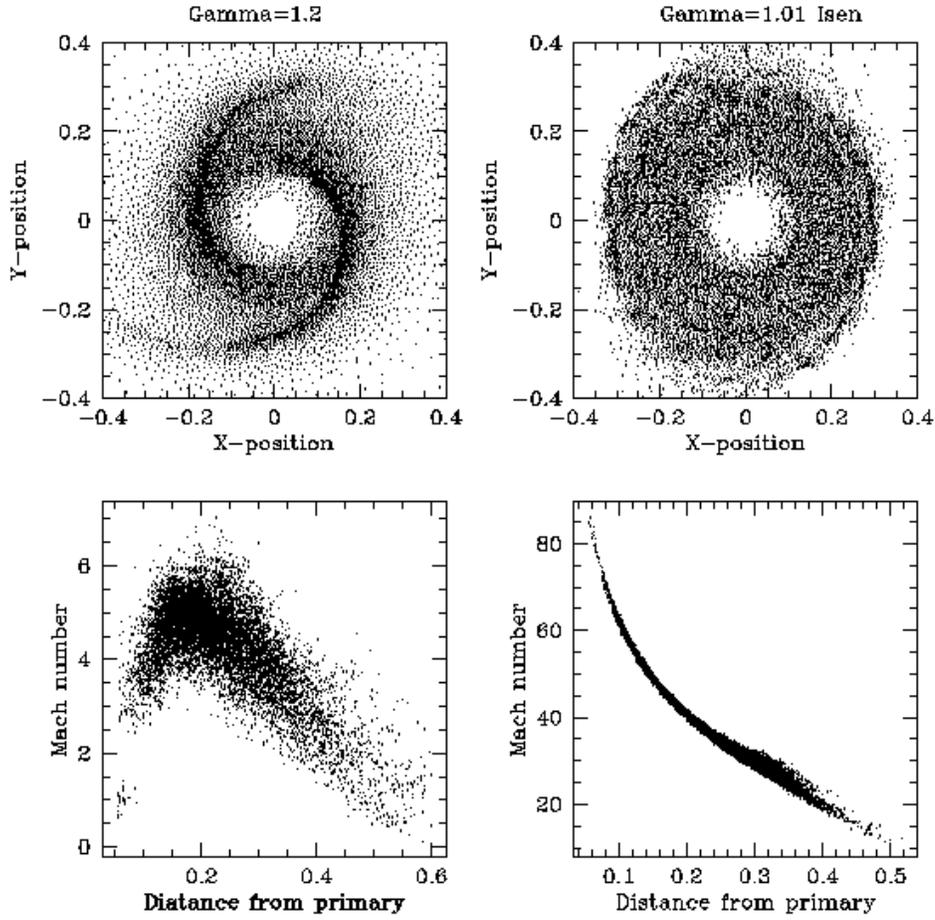}
\caption{
\label{fig:comp}
Results of 2D numerical simulations using SPH. The upper graphs
show the flow in the orbital plane, while the lower ones show the Mach number as
a function of the distance to the primary. Left: A polytropic eos with $\gamma=1.2$.
Right: A pseudo-isentropic eos $\gamma=1.01$}
\end{figure}

\begin{figure}
\begin{center}
\epsfig{file=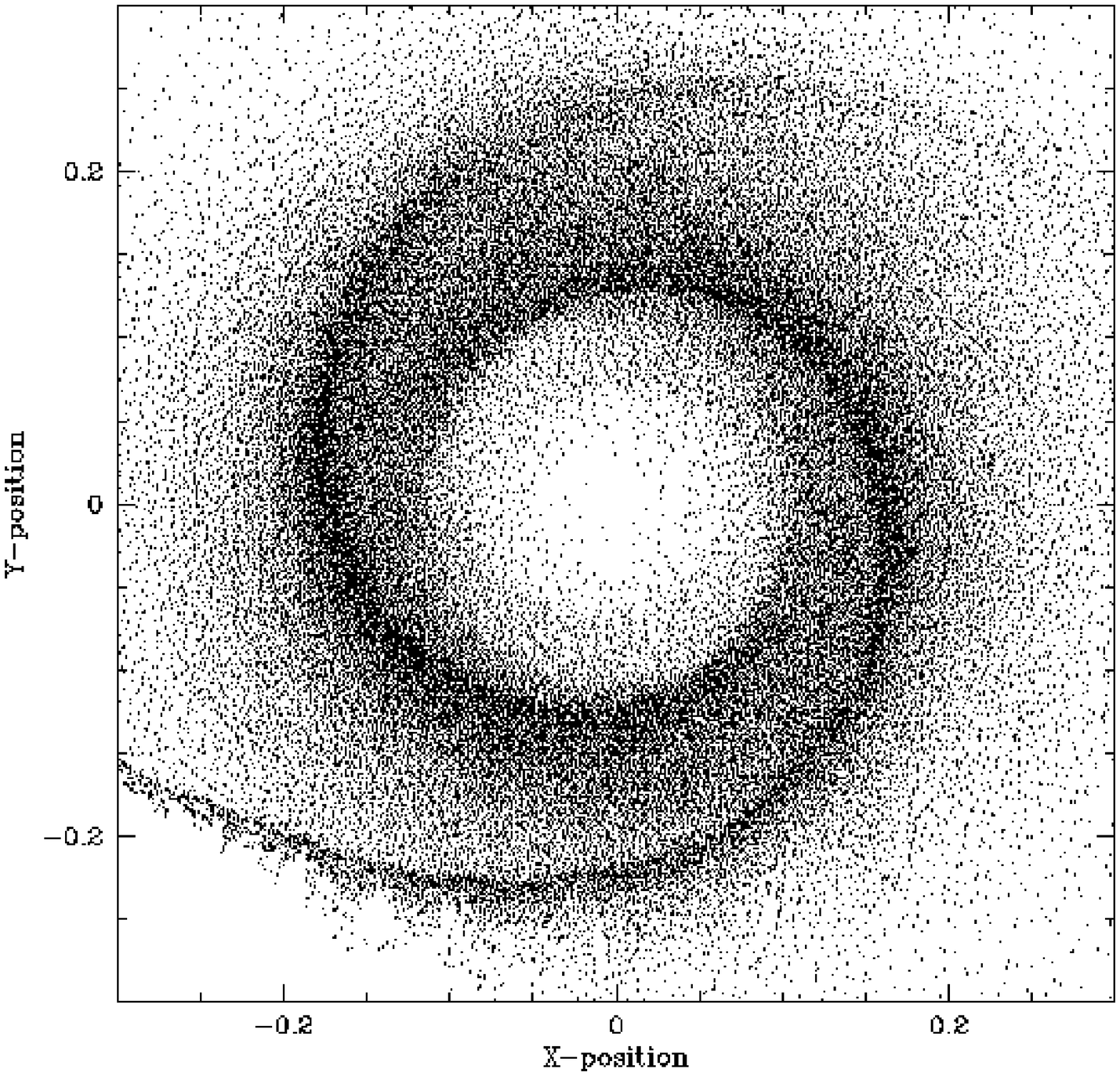,height=8cm}
\end{center}
\caption{
\label{fig:nsph12}
Structure of the flow in the orbital plane for a polytropic eos with $\gamma=1.2$
as obtained with a modified SPH code.}
\end{figure}
\section{Theoretical analysis}
Spirals are an ubiquitous phenomenon in astrophysical sites. They may appear, for example,
in galaxies, protostellar discs, associated with Jovian planets in proto-planetary discs
as well as in accretion discs in binary systems.
Such waves may play a very crucial role in the transport of angular momentum and hence in 
the process of accretion.

\begin{figure}
\begin{center}
\epsfig{file=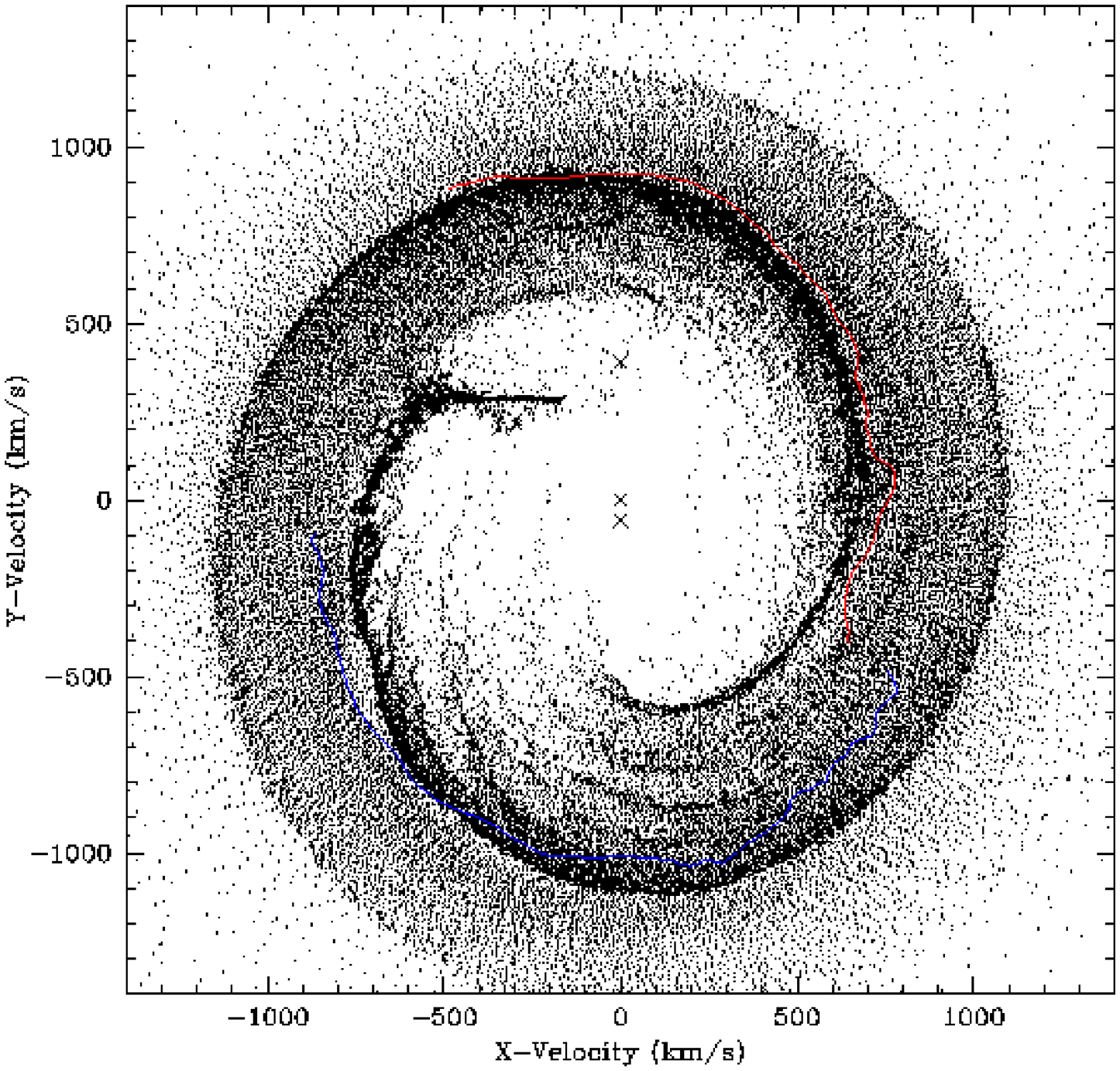,height=8cm}
\end{center}
\caption{
\label{fig:tomonsph12}
Doppler map corresponding to the flow shown in Fig.~3. Also plotted as solid lines 
are the location of the spirals as observed in the August 1999 outburst of IP Peg.}
\end{figure}

Tidally generated spiral waves are the result  of tidal torques of the
companion star on  the orbiting disc  material. Initially triggered at
the outer edge  of the disc, where  the tidal interaction between  the
disc and   the companion star  is   strongest, they take  the  form of
trailing  spirals because the azimuthal  velocity of the disc material 
monotonically increases inwards.
Lin \& Papaloizou (1979) were among the first to observe the spiral pattern
in their study of tidal torques on accretion discs in binary systems
with extreme mass ratios. In their case, the spirals were the result of the interaction
with the inner Lindblad 2:1 resonance. It was however Sawada, Matsuda, \& Hachisu (1986a,b)
who showed, in their 2D inviscid numerical simulations of accretion discs in a binary of unit mass ratio, that spiral shocks
could form which propagate to very small radii. Spruit (1987) and later Larson (1990) made semi-analytical calculations which were followed by numerous - mostly 2D - numerical simulations. A more detailed account can be found in Boffin (2001).

Savonije, Papaloizou, \& Lin (1994) presented both linear and non-linear
calculations of the tidal interaction of an accretion disc in close binary systems.
They showed, in agreement with Lin \& Papaloizou (1979), that the resonance is significant
over a region proportional to the temperature. 
Thus, even in CVs with larger mass ratios, they concluded that the centre of the 2:1 resonance can still be thought of as
lying in the vicinity of the boundaries of the disc and that  
the resonance can
still generate a substantial wave-like spiral response in the disc, but only if
the disc is large and the Mach number is smaller than about 10.  For
larger Mach number, Savonije et al. consider that wave excitation and propagation becomes ineffective and
unable to reach small radii at significant amplitude.

\subsection{Opening angle}
The opening angle of the spirals is directly related to the temperature of the disc as,
when the shock is only of moderate strength, it roughly propagates at sound speed.
For illustration purposes, we present in Fig.~\ref{fig:comp} 
two numerical simulations of an accretion disc using the Smoothed Particle Hydrodynamics (SPH) method.
An adiabatic equation of state (eos) is used with $\gamma=1.2$ in one case, and a pseudo-isentropic 
eos with $\gamma=1.01$ in the other. As is clear from the figure, the $\gamma=1.2$ case produces a very open 
spiral, while the cooler one shows only a hint of disturbance 
at the outer edge, with very tightly wound spirals. This is even more clear in the high
resolution simulation shown in Fig.~7 of Boffin (2001). Figure~\ref{fig:comp} also shows the reason for
the difference: in the former case, the Mach number has values as small as 5, corresponding to a very hot disc, while in
the latter case, the Mach number is above 20. This effect is the principal reason why it was believed that spirals may 
not be present in the accretion discs of cataclysmic variables, which are not hot enough.
This is in particular the case in quiescent discs.
As a byproduct, Fig.~\ref{fig:comp} also shows that the origin of the spirals lies in the tidal force of the companion, as in these simulations, no mass transfer is allowed from the companion.

\subsection{Structure of spiral shocks}
\begin{figure}
\begin{center}
\epsfig{file=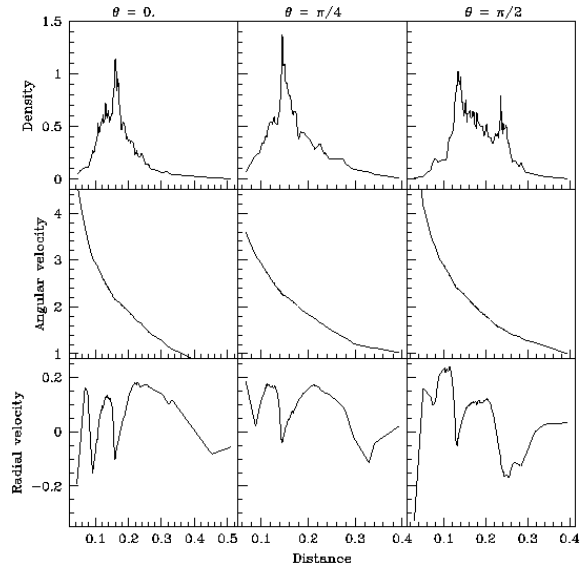,height=8cm}
\end{center}
\caption{
\label{fig:slice}
Slices through the flow shown in Fig.~3 at 3 angles: $0, \pi/4$ and $\pi/2$.
The density, angular and radial velocity are shown. The position of the spiral shock is obvious.}
\end{figure}
In this section, we would like to have a more detailed look at the way spirals affect the flow and hence lead to 
angular momentum transfer and mass accretion. To this aim, we will use 
the result of a simulation with $\gamma=1.2$ using a modified version of the SPH
method (Boffin, Yukawa, \& Matsuda, in preparation). Figure~3 shows the flow in
a close binary with mass ratio $q=0.15$ when we allow mass transfer from the inner Lagrangian point.
Again, the now very typical open spiral can be seen.
We then make three slices in this flow, at angles 0, $\pi/4$ and $\pi/2$, considering only the
upper right quadrant, and show the properties of the particles as a function of distance from a primary star
 (Fig.~\ref{fig:slice}).
As can be seen, the spiral shocks have a clear effect on the density field but also on the velocity
field. In particular, the radial velocity exhibits a clear jump at the position of the shock. A more detailed analysis of the role of the spirals
in the transport of angular momentum, in terms of an effective $\alpha$ 
can be found in Boffin (2001).
In Fig.~\ref{fig:tomonsph12} we show the corresponding Doppler map, that is the structure of 
the flow in the velocity-velocity plane. This allows a direct comparison with observations.

\subsection{Comparison with observations}
Looking at Fig.~\ref{fig:tomonsph12} and \ref{fig:ugem}, one sees that a qualitative agreement
can easily be reached. In fact, in Fig.~\ref{fig:tomonsph12}, we have also plotted a scaled version of the
position and intensity of the spirals observed during the August 1999 outburst of IP Peg
(Steeghs \& Boffin, in preparation; see Steeghs 2001). The agreement between theory and observations seems rather 
convincing. This very fact is at the base of a lot of controversy in the literature, because the discs obtained 
in the numerical simulations are considered too hot to be realistic, even for systems in outburst (see e.g. Savonije et al.
1994, Godon, Livio, \& Lubow 1998). 

But what is the Mach number in the discs of cataclysmic variables ?
A rough estimate is generally taken assuming Keplerian rotation and a central temperature of a few $10^4$ K. This gives
then Mach numbers in the range 15 to 50. One has to note however that, because of the tidal field, the flow
in the accretion disc is far from being Keplerian, and large departures exist that can affect the geometry of the spirals, especially in the 
outer parts. Moreover, the tidal force can increase the temperature of the these outer parts (Steeghs \& Stehle 1999).
These two effects join together to bring the Mach number to values smaller than about 10, at least in 
the outer parts of the disc. This could explain the apparent agreement between observations and theory.
This explains also why spirals can only be seen during the outburst, when the disc is large and hot. However, it has still to be explored if other parameters
than the Mach number do not play a role in the geometry of the spiral pattern.
A definitive conclusion will only be possible once we can make detailed three-dimensional numerical simulations
including a realistic equation of state. One should also caution that the spirals observed in the different 
cataclysmic variables are not always similar and that even for a given system, their shape may vary in the 
outburst cycle.
As Groot (2001) noticed, 
its tomogram of U Gem  nine days after maximum compares favorably with the low Mach number simulations in 
Steeghs \& Stehle (1999), while the IP Peg, EX Dra and U Gem in decline tomograms, all appear to have their
maximum emissivity rotated anti-clock wise with respect to the
simulations of Steeghs \& Stehle (1999). On the other hand, as
shown by Fig.~\ref{fig:tomonsph12}, it is also possible to have an agreement between numerical models and one outburst of IP Peg.
Here again, one needs detailed self-consistent numerical simulations and a complete observational coverage of 
the same outburst of one cataclysmic variable to reach a more definitive conclusion. The time of pioneering is over, now comes the time for a
detailed study.

\end{document}